\documentclass[%
 reprint,
showpacs,
 amsmath,amssymb,
 aps,
 prb,
 twocolumn,
floatfix,
showkeys,
]{revtex4-1}

\usepackage{color}
\usepackage{graphicx}% Include figure files
\usepackage{dcolumn}% Align table columns on decimal point
\usepackage{bm}% bold math
\usepackage{hyperref}% add hypertext capabilities
\definecolor{color10}{rgb}{1,0,0}
\definecolor{color12}{rgb}{0.75,0,0.25}
\definecolor{color14}{rgb}{0.5,0,0.5}
\definecolor{color16}{rgb}{0.25,0,0.75}
\definecolor{color18}{rgb}{0,0,1}

\definecolor{coloro10}{rgb}{1,0.5,0}
\definecolor{coloro12}{rgb}{0.9,0.45,0}
\definecolor{coloro14}{rgb}{0.8,0.4,0}
\definecolor{coloro16}{rgb}{0.7,0.35,0}
\definecolor{coloro18}{rgb}{0.6,0,0}

\definecolor{colorb10}{rgb}{0,0.5,1}
\definecolor{colorb12}{rgb}{0,0.4,0.8}
\definecolor{colorb14}{rgb}{0,0.3,0.6}
\definecolor{colorb16}{rgb}{0,0.2,0.4}
\definecolor{colorb18}{rgb}{0,0.1,0.2}

\newcommand{\ket}[1]{\ensuremath{\left| #1 \left\rangle \right. \right.}}

\newcommand{\elemm}[3]{\ensuremath{\left\langle #1 \left| #2 \right| #3 \right\rangle}}

\begin{document}

\preprint{APS/123-QED}

\title{First-principles study of excitonic effects in Raman intensities.}% Force line breaks with \\

\author{Yannick Gillet}
 \email{yannick.gillet@uclouvain.be}
\author{Matteo Giantomassi}%
\author{Xavier Gonze}
\affiliation{% 
 Universit\'e catholique de Louvain, 
 Institute of Condensed Matter and Nanosciences, 
 Nanoscopic Physics,
 Chemin des \'etoiles 8, bte L7.03.01, 
 1348 Louvain-la-Neuve, Belgium
}%

\date{\today}% It is always \today, today,

\begin{abstract}
The ab initio prediction of Raman intensities for bulk solids usually relies on the hypothesis that
the frequency of the incident laser light is much smaller than the band gap.
However, when the photon frequency is a sizeable fraction of the energy gap, or higher, resonance effects appear.
In the case of silicon, when excitonic effects are neglected, 
the response of the solid to light increases by nearly three orders of magnitude in the range of frequencies between the static limit and the gap. When excitonic effects are taken into account, 
an additional tenfold increase in the intensity is observed.  We include these effects using a finite-difference scheme applied on the dielectric function obtained by solving the Bethe-Salpeter equation.  
Our results for the Raman susceptibility of silicon show stronger agreement with experimental data compared with previous theoretical studies. 
For the sampling of the Brillouin zone, a double-grid technique is proposed, resulting in a significant reduction in computational effort. 
\end{abstract}


%Use showkeys class option if keyword
\maketitle

%\tableofcontents

\section{Introduction}

Raman spectroscopy is widely used to characterize materials by means of their vibrational fingerprint.
The dependence of the Raman intensity on the frequency of the incident light is well-known. It is, for example, used to amplify the Raman response, resulting in the appearance of a resonance phenomenon when the frequency of the exciting light is close to electronic transitions.\cite{Weber2000}
Unlike for molecules\cite{Stock1995,Dresselhaus2005} and for graphene,\cite{Malard2009,Venezuela2011} the first-principles prediction of the frequency dependence of the Raman intensity of crystalline systems has received little attention.\cite{Weber2000}

The Raman intensity is related to the derivative of the macroscopic dielectric function with respect to collective atomic displacements. 
Different first-principle formalisms have been proposed for the computation of such a dielectric function. These formalisms often trade computational speed for predictive power, or vice versa. In the present study, a method that provides an accurate description of the dielectric properties
 of material was chosen in order to establish the importance of different physical effects, and in particular, excitonic effects.

Within the static limit (vanishing light frequency), the dielectric response can be computed with Density-Functional Theory (DFT)\cite{Hohenberg1964,Kohn1965,Martin2004} followed by Density-Functional Perturbation Theory (DFPT).\cite{Baroni1987, Gonze1997a, Baroni2001} Although DFT is plagued by the well-known band gap problem,\cite{Martin2004} its prediction of the static dielectric tensor is reasonably accurate (to within 5-10\%) except when the gap is very small.\footnote{A true semiconductor might be described as a metal by DFT {\it e.g.} with usual local or semi-local functionals\cite{Godby1989,Ghosez1997}}
Subsequent computation of the derivative of the dielectric tensor with respect to an atomic displacement can be performed by using either finite differences~\cite{Baroni1986} or the $2n+1$ theorem of perturbation theory.\cite{Veithen2004,Veithen2005} Such methodology
has been applied in numerous studies.\cite{Hermet2012,Zhang2011} As an example, more than two hundred Raman spectra are provided in the WURM database.\cite{CaracasWURM, Caracas2011}

When the excitation frequency is comparable to the gap, DFT becomes unreliable for the prediction of the dielectric response. 
Not only does the proximity of the resonance increase the need to rely on an accurate band gap, but excitonic effects also drastically modify the optical properties of most semiconductors.\cite{Yu2010} The band-gap correction is usually treated within the $GW$ approximation of Many-Body Perturbation Theory (MBPT), while
the Bethe-Salpeter Equation (BSE) is the method of choice to introduce excitonic effects.\cite{Aulbur1999, Onida2002}
To our knowledge, the BSE has not yet been used to compute Raman intensities of solids.  The purpose of the present work is to compute the Raman intensities, using a finite difference approach that combines multiple BSE results performed for different atomic displacements. 

Excitonic effects can also be addressed within the framework of time-dependent density functional theory.\cite{Runge1984,Onida2002, Botti2004, Marques2006} This approach, which is computationally cheaper, also allows one to include excitonic effects, with an accuracy that depends on the choice of the exchange-correlation kernel. 
Recent studies shows interesting agreement with experiment for the macroscopic dielectric function (see e.g. Ref. \onlinecite{Sharma2011}). This route is not pursued in the present study. Instead we rely on the best theoretical approach available today to compute the frequency-dependent Raman intensities, and examine its predictive power in comparison with experimental data.

We chose to study silicon, for which the experimental frequency-dependent enhancement factor is particularly strong. 
The available data cover the frequency range between 1.8 eV and 3.8 eV,\cite{Compaan1984} and the experimental value of the direct gap is at 3.4 eV. Due to the high symmetry of silicon, there is only one Raman-active phonon mode, whose eigenvector is determined by symmetry. 

% The Raman experimental measurements provide the ratio between the Raman intensity of silicon and the intensity of a reference material. This procedure is used in order to benefit from cancellation of (less relevant) frequency-dependent factors. CaF$_2$ is chosen in Ref.~\onlinecite{Renucci1975}. 
% Thus, we also compute the frequency-dependent Raman intensity of CaF$_2$, in the frequency range that is relevant for silicon (below the CaF$_2$ band gap).
% Note that also in CaF$_2$, due to the high symmetry of the crystal, there is only one Raman-active phonon mode whose eigenvector is determined by symmetry. 

In Sec. II of this article, the theoretical basis needed for the computation of the resonant Raman intensities is described, taking into consideration the main equations of the MBPT in the GW and BSE frameworks. 
Sec.~III describes the numerical procedure.
In Sec. IV, the problem associated with the slow convergence of results with respect to the sampling of the Brillouin zone is analyzed.
Sec. V presents the theoretical results, including excitonic effects.
Finally, in Sec. VI, theoretical and experimental results for the silicon Raman intensity are compared.

\section{The computation of resonant Raman intensities for solids}

The scattering efficiency (time-average of the power radiated into unit solid angle) of the phonon of frequency~$\omega_m$ for a photon of frequency~$\omega_i$ is defined as:\cite{Veithen2005}
\begin{eqnarray}
 I &=& (\omega_i - \omega_m)^4 | \textbf{e}_o . \boldsymbol{\alpha}^m . \textbf{e}_i |^2 \frac{n_m +1}{2 \omega_m}, \label{eq:Iquasi}
\end{eqnarray}
with $\textbf{e}_o$ and $\textbf{e}_i$ the outgoing and ingoing polarization of the light, $n_m$ the phonon occupation factor:

\begin{equation}
 n_m = \frac{1}{e^{{\omega_m}/{kT}}-1}.
\end{equation}
The complete field-theoretic expression for the Raman susceptibility $\alpha^m(\omega)$ is presented in Ref \onlinecite{Yu2010}.
It includes six terms, in which the frequencies $\omega_m$ and $\omega_i$ are combined in different denominators,  giving resonant as well as anti-resonant contributions. 
In the following calculations, we will use the quasi-static approximation which neglects the dynamical effects due to the phonons. 
Mathematically, this approximation is well-justified\cite{Renucci1975} when:
\begin{equation}
 \omega_m \ll \left| \omega_i - \omega_{gap} + i\eta \right|
\end{equation}

with $\omega_{gap}$ the frequency corresponding to the direct band gap and $\eta$ the lifetime broadening of the gap.

In this framework, the Raman susceptibility $\boldsymbol{\alpha}^m(\omega)$ for the phonon $m$ is defined as:

\begin{equation}
 \alpha^m_{ij}(\omega) = \sqrt{\Omega_0} \sum_{\tau \beta} \frac{\partial \chi_{ij}(\omega)}{\partial R_{\tau \beta}} u^m_{\tau \beta} \label{ramantensor}
\end{equation}
with $\Omega_0$ the unit cell volume, $\chi_{ij}$ the macroscopic dielectric susceptibility and $u^m_{\tau\beta}$ the eigendisplacement of phonon mode $m$ of atom $\tau$ in direction $\beta$. 
In the present work, we neglect higher-order derivatives with respect to atomic displacements.
The eigendisplacements are normalized as:

\begin{equation}
 \sum_{\tau,\beta} M_\tau u^m_{\tau \beta} u^n_{\tau \beta} = \delta_{mn},
\end{equation}
with $M_\tau$ the mass of atom $\tau$.\footnote{Atomic units are used through all this paper. 
One atomic unit for $\alpha$ corresponds to $7.62~10^9~\text{m}^{1/2}/\text{kg}^{1/2}$.}
For more details about the derivation of Eq. \eqref{eq:Iquasi}, we refer the reader to Refs. \onlinecite{Veithen2005} and \onlinecite{Renucci1975}.

We define the Raman polarizability $a$ by:
\begin{equation}
 a = \sqrt{\mu \Omega_0} \alpha
\end{equation}
with $\mu$ the reduced mass (in the case of silicon, $\mu~=~M_{Si}/2$).

When the incoming frequency $\omega_i$ is close to the
energy of the direct gap, there is a resonant process and the amplitude of $\boldsymbol{\chi}(\omega_i)$  and $\boldsymbol{\alpha}^m(\omega_i)$ 
can change by several orders of magnitude. 

The computation of the macroscopic dielectric susceptibility $\boldsymbol{\chi}(\omega_i)$ follows the standard procedure used in ab initio MBPT. 
Two steps are needed: the computation of the quasiparticle energies, followed by the computation of the dielectric response of the material.
The quasiparticle amplitudes $\psi_{i}^{\mathrm{QP}}$ and the quasiparticle energies $\epsilon_{i}^{\mathrm{QP}}$ are computed by solving the following equation:

\begin{multline}
\left(-\frac{1}{2}\nabla^2 + V_{\mathrm{ext}}(\mathbf{r})+V_{\mathrm{H}}(\mathbf{r})\right)\psi_{i}^{\mathrm{QP}}(\mathbf{r}) \\
+ \int d\mathbf{r'} \Sigma(\mathbf{r},\mathbf{r'};\omega=\epsilon_{i}^{\mathrm{QP}})\psi_{i}^{\tiny{\mathrm{QP}}}(\mathbf{r}') = \epsilon_{i}^{\mathrm{QP}} \psi_{i}^{\mathrm{QP}}(\mathbf{r}), \label{eqn:quasiparticle_equation}
\end{multline}

with $V_{\mathrm{ext}}(\mathbf{r})$ the electrostatic potential of the ions and $V_{\mathrm{H}}(\mathbf{r})$ the Hartree potential originating from the electronic density $n(\mathbf{r})$.
In Eq. \eqref{eqn:quasiparticle_equation}, $\Sigma(\mathbf{r},\mathbf{r'};\omega)$ is the self-energy, that, in the so-called GW approximation, is given by:

\begin{equation}\label{eqn:sigma_equation}
\Sigma(\mathbf{r},\mathbf{r'};\omega)=\frac{i}{2\pi}\int d\omega' G(\mathbf{r},\mathbf{r'},\omega+\omega')W(\mathbf{r},\mathbf{r'},\omega')
\end{equation}
where $G$ is the Green's function and $W$ the screened Coulomb interaction.\cite{Onida2002}

In the second part of the calculation, we include excitonic effects by working within the BSE framework.\cite{Onida2002}
In this framework, we introduce $H$, a two-particle hamiltonian that describes the interaction between electrons and holes.
In the transition space, formed by products of two Kohn-Sham orbitals, the BSE hamiltonian has the following block structure:

\begin{equation}
 H = \begin{pmatrix}
      & \ket{v'c'\textbf{k}'} & \ket{c'v'\textbf{k}'} \\
      \ket{vc\textbf{k}} & R & C \\
      \ket{cv\textbf{k}} & -C^* & -R^*
     \end{pmatrix}
\end{equation}
where $v$, $c$ and $\textbf{k}$ denotes the valence band index, the conduction band index and the wavevector. 

The resonant sub-block $R$ is Hermitian, and the coupling term $C$ is symmetric. 
Due to the coupling sub-blocks that connect resonant and anti-resonant transitions, the Bethe-Salpeter Hamiltonian is not Hermitian. This complicates the solution of the problem.
In crystalline systems, however, the matrix elements of $C$ are usually much smaller than the matrix elements of $R$.
For this reason, the matrix elements of $C$ are usually neglected when solving the Bethe-Salpeter problem in extended systems -- the so called Tamm-Dancoff approximation (TDA).\cite{Fetter1971}
This approximation is used in all the rest of this work.

The matrix elements of the resonant block are given by:

\begin{equation}
 R_{(vc\textbf{k}),(v'c'\textbf{k}')} = H^{diag}_{(vc\textbf{k}),(v'c'\textbf{k}')} + H^{exch,R}_{(vc\textbf{k}),(v'c'\textbf{k}')} + H^{coul,R}_{(vc\textbf{k}),(v'c'\textbf{k}')}
\end{equation}
where

\begin{widetext}
\begin{align}
 H^{diag}_{(vc\textbf{k}),(v'c'\textbf{k}')} &= (\epsilon_{c\textbf{k}} - \epsilon_{v\textbf{k}}) \delta_{vv'} \delta_{cc'} \delta_{\textbf{k}\textbf{k}'} \label{eq:Hdiag} \\
 H^{exch,R}_{(vc\textbf{k}),(v'c'\textbf{k}')} &= 2 \elemm{vc\textbf{k}}{\bar{v}}{v'c'\textbf{k}} = 2 \int \int \psi_{v\textbf{k}}(\textbf{r}) \psi_{c\textbf{k}}^*(\textbf{r}) \bar{v}(\textbf{r} - \textbf{r}') \psi_{v'\textbf{k}'}^*(\textbf{r}') \psi_{c'\textbf{k}'}(\textbf{r}') d\textbf{r}' d\textbf{r} \label{eq:Hexch} \\
 H^{coul,R}_{(vc\textbf{k}),(v'c'\textbf{k}')}&= - \elemm{vc\textbf{k}}{W}{v'c'\textbf{k}'} = - \int \int \psi_{v\textbf{k}}(\textbf{r}) \psi_{v'\textbf{k}'}^*(\textbf{r}) W(\textbf{r},\textbf{r}') \psi_{c\textbf{k}}^*(\textbf{r}') \psi_{c'\textbf{k}'}(\textbf{r}') d\textbf{r}' d\textbf{r} \label{eq:Hcoul}
\end{align}
\end{widetext}
with $\bar{v}$ the modified Coulomb potential, whose Fourier transform does not contain the $\textbf{q} = 0$ component:
\begin{align}
 \bar{v}(\textbf{q}) = \begin{cases}
			   v(\textbf{q}) &\mbox{ if } \textbf{q} \ne 0 \\
			   0 &\mbox{ if } \textbf{q} = 0,
			  \end{cases}
\end{align}
$v(\textbf{r})$ the standard Coulomb potential:
\begin{align}
 v(\textbf{r}) = \frac{1}{|\textbf{r}|},
\end{align}
$W(\textbf{r},\textbf{r}')$ the screened Coulomb potential:
\begin{align}
 W(\textbf{r},\textbf{r}') = \int d\textbf{r}'' \epsilon^{-1}(\textbf{r},\textbf{r}'') v(\textbf{r}''-\textbf{r}'), \label{eq:Wcoul}
\end{align}
and $\epsilon^{-1}(\textbf{r},\textbf{r}')$ the inverse dielectric function.
For the derivation of Eqs. \eqref{eq:Hdiag} to \eqref{eq:Wcoul}, we refer to Ref. \onlinecite{Onida2002}.

The dielectric susceptibility $\chi(\omega)$ and macroscopic dielectric function $\varepsilon(\omega)$ are then obtained from:

\begin{eqnarray}\label{mdf}
 \varepsilon(\omega) &=& 1 + 4 \pi \chi(\omega)
\\
&=&  1 - \lim_{\textbf{q} \rightarrow 0} v(\textbf{q}) \elemm{P(\textbf{q})}{((\omega + i\eta) - H)^{-1} F}{P(\textbf{q})} 
\nonumber
\\
\end{eqnarray}
where $\eta$ is a broadening factor, $F$ is taking into account the occupation numbers:

\begin{equation}
 F = \begin{pmatrix}
      & \ket{v'c'\textbf{k}'} & \ket{c'v'\textbf{k}'} \\
      \ket{vc\textbf{k}} & 1 & 0 \\
      \ket{cv\textbf{k}} & 0 & -1
     \end{pmatrix}
\end{equation}
and

\begin{equation}
 P(\textbf{q})_{(n_1n_2)} = \elemm{n_2}{e^{i\textbf{q}.\textbf{r}}}{n_1},
\end{equation}
are the so-called oscillator matrix elements where $n_1$ and $n_2$ are a short-hand notation for $vc\textbf{k}$.

The Random-Phase approximation (RPA) is a simplification of the BSE approach,
in which the exchange\footnote{The exchange contributions, Eq. \eqref{eq:Hexch} are often referred to as being local-field effects.} and Coulomb terms, Eqs. \eqref{eq:Hexch} and \eqref{eq:Hcoul}, are neglected. In the RPA, the BSE Hamiltonian $H$ is diagonal, and the spectrum
is obtained directly as a simple sum over transitions between valence and conduction bands, weighted by the proper oscillator matrix elements. In such an independent-particle approximation, no excitonic effect is present. The importance of the excitonic
effect on the optical spectrum is well-known, 
with prominent peaks being created below the band gaps in most wide-gap insulators or semiconductors,
and redistribution of the spectral weight.

\section{Numerical procedure}

Calculations are performed using ABINIT.\cite{Gonze2005,Gonze2009}
The pseudopotential used to simulate the silicon atom is of the Troullier-Martins type used in the Teter parametrization.
% The pseudopotential described in Ref.~\onlinecite{Ponce2013} is used to reproduce accurately the core states of the calcium atoms for CaF$_2$. 
% For the fluorine, the pseudopotential is of Troullier-Martins type, with a Perdew-Wang, Ceperley/Alder parametrization.

The DFT-LDA calculations are performed with a 4 times shifted 4x4x4 Monkhorst-Pack grid to sample the Brillouin Zone (BZ),\cite{Monkhorst1976} and a plane-wave basis set kinetic energy cut-off of 16 Ha.
The theoretical lattice cell for silicon is 10.20 Bohr, which gives an error of 0.6 \% with respect to the experimental results (10.26 Bohr).\cite{Wyckoff1963}
Using this theoretical lattice constant, the DFT-LDA indirect gap is 0.45 eV, while the direct gap is 2.52 eV. 
 
Quasi-particle corrections are computed within the so-called \textit{one-shot} $GW$ or $G_0W_0$ approximation.\cite{Aulbur1999}
We use a cut-off energy of 8~Ha for the screening and 16 Ha for the self-energy matrix elements.
An extrapolar energy of 3~Ha is used to reduce the number of bands needed to converge to 100 bands.\cite{Bruneval2008}
The computed GW corrections give a direct gap of 3.20~eV. 
These results are comparable to other GW results\cite{Ku2002, Arnaud2000} and in good agreement with the experimental band gap of 3.4 eV.\cite{Landolt-Bornstein1982}
During the computation of the BSE optical spectrum, the opening of the gap is simulated by a rigid scissor\cite{Levine1989} with a value of 0.65 eV to reproduce the theoretical GW gap unless stated otherwise.

Convergence of the Bethe-Salpeter computations with respect to the BZ sampling is particularly difficult, and is discussed in the next section.
The cut-off energies are 16 Ha for the wavefunctions and 3~Ha for the screening. 
The included bands range from the second to the ninth band.
A broadening factor of 0.1 eV is used for the dielectric function.

% For CaF$_2$, the relaxation of the unit cell gives a theoretical lattice parameter of 10.07~Bohr, which is used throughout the calculations. 
% This value is in close agreement with the value reported in Refs. \onlinecite{Verstraete2003,Ma2007}. 
% The experimental lattice parameter is 10.32~Bohr, with a relative error of 2.4\% with respect to the computed value.
% The LDA direct gap is 7.6~eV which is 4.5~eV lower than the experimental gap of 12.1~eV.\cite{Rubloff1972} 
% A rigid shift of 4.5~eV is used to simulate the opening of the gap.
% The cut-off energies used to obtain converged Raman intensities are 12~Ha for the screening and 65~Ha for the wavefunctions. 
% The 100 first bands are included to compute the screening. Bands from index 4 to index 30 are included as the basis set for the BSE part.
% 
% The evolution of the Raman intensity of the unique Raman peak of silicon and CaF$_2$ with respect to the frequency is obtained by finite differences over multiple atomic displacements (frozen-phonon approach). 
% The symmetries of these crystals allow only one (triply degenerate) Raman-active mode.
% In the CaF$_2$, the Raman-active mode involves an out-of-phase displacement of the two fluorine atoms.

The quasi-static approximation that is used extensively in this work is justified in the case of silicon since the lifetime broadening ($\approx 0.1 \text{ eV}$) is larger than the phonon frequency $\approx 0.065 \text{ eV}$.\cite{Renucci1975}

In this quasi-static approximation, two-band as well as three-band contributions to the resonant Raman are included, the latter coming from the matrix element changes due to changes in the wavefunctions produced by phonon-induced admixture of the two bands under consideration with a third band.\cite{Renucci1974}

In order to compute derivatives with the displacements, we add $h \times \sqrt{2}/{2}$ to the x-position of the atom and $-h \times \sqrt{2}/{2}$ to the x-position of the other atom, for different values of $h$.
The derivative is obtained by computing $\chi_{yz}(\omega)$ for $h = 0.01$ and $h = 0$ in the convergence studies and is obtained by computing $\chi_{yz}(\omega)$ for $h = 0.01$ and $h = -0.01$ for the final result.

We have analyzed the behaviour of the G$_0$W$_0$ scissor shift, as a function of the
frozen phonon amplitude. Because the Raman amplitude corresponds to a first-order
derivative with respect to atomic displacement, see Eq.~\eqref{ramantensor}, we only have to consider the linear response.
For non-degenerate eigenstates, due to the high symmetry of the crystals, 
such derivative of the scissor shift
with respect to atomic positions vanish. For degenerate eigenenergies, linear variations of eigenenergies are present,
but the mean variation vanishes over the set of degenerate states. Hence, the modification of the G$_0$W$_0$ corrections
with respect to atomic displacement does not have any effect on the Raman intensity of silicon,
within the present formalism. Of course, this is a very specific situation. For the analysis of most other materials,
the variation of the eigenenergies at linear order will have to be taken into account.

As implemented in ABINIT, the BSE gives the macroscopic dielectric function for a given $q$-direction:

\begin{align}
 \varepsilon(\omega,\textbf{q}) = \frac{\textbf{q}^T \boldsymbol{\hat{\varepsilon}} \textbf{q}}{\textbf{q}^T \textbf{q}}
\end{align}
where $\boldsymbol{\hat{\varepsilon}}$ is the dielectric tensor.

Because of the symmetry of the tensor, the values of $\boldsymbol{\hat{\varepsilon}}$ for each pair of cartesian coordinates may be obtained by computing the macroscopic dielectric function for 6 different directions $q$.
% Because of the symmetry of the tensor, computing this value for 6 different directions $q$ allows to obtain the values of $\boldsymbol{\hat{\varepsilon}}$ for each pair of cartesian coordinates.
The directions used in the calculations are (0,1,1), (1,0,1), (1,1,0), (1,0,0), (0,1,0) and (0,0,1).

The macroscopic dielectric function (Eq. \eqref{mdf}) is computed using the iterative Haydock technique.\cite{Haydock1980} The algorithm is terminated when a relative error of 1\%, for the real and the imaginary part of $\varepsilon$, is achieved for each frequency in the frequency range under investigation.

\section{Sampling of the Brillouin zone}

As mentioned previously, achieving converged results with respect to the sampling of the Brillouin zone is a very difficult issue.
In particular, we will show in this section that grids that are appropriate for obtaining a converged 
macroscopic dielectric function in the whole frequency range are not sufficiently dense for derivatives,
such as the Raman intensities, at least in the resonance region.

In order to accelerate the convergence of BSE spectra, shifted homogeneous meshes are traditionally used.
A symmetry-breaking shift allows one to sample more non-equivalent points and therefore leads to a more representative sampling of the band dispersion: with respect to non-shifted meshes, its presence lowers the computational load for an equivalent convergence criterion.

To ease the discussion, we introduce specific notations.
The meshes are characterized by the number of divisions along each axis of the primitive cell in reciprocal space, namely $n_1$, $n_2$ and $n_3$. 
For a crystalline cubic structure these three numbers are taken as equal ($n_1~=~n_2~=~n_3~=~n_k$).
The total number of points inside this mesh is therefore $N_k = n_1 n_2 n_3 = n_k^3$.
All the points of the mesh are shifted by a certain vector
characterized by three real numbers $s_i$ between -0.5 and 0.5,
$\textbf{s} = (s_1,s_2,s_3)$.
This shift is such that the point $(s_1/n_1,s_2/n_2,s_3/n_3)$ belongs to the shifted mesh.
We use the notation $(n_1 \times n_2 \times n_3| \textbf{s} )$ to refer to such a mesh.

\begin{figure}
 \def\svgwidth{8cm}
\begingroup%
  \makeatletter%
  \providecommand\color[2][]{%
    \errmessage{(Inkscape) Color is used for the text in Inkscape, but the package 'color.sty' is not loaded}%
    \renewcommand\color[2][]{}%
  }%
  \providecommand\transparent[1]{%
    \errmessage{(Inkscape) Transparency is used (non-zero) for the text in Inkscape, but the package 'transparent.sty' is not loaded}%
    \renewcommand\transparent[1]{}%
  }%
  \providecommand\rotatebox[2]{#2}%
  \ifx\svgwidth\undefined%
    \setlength{\unitlength}{448bp}%
    \ifx\svgscale\undefined%
      \relax%
    \else%
      \setlength{\unitlength}{\unitlength * \real{\svgscale}}%
    \fi%
  \else%
    \setlength{\unitlength}{\svgwidth}%
  \fi%
  \global\let\svgwidth\undefined%
  \global\let\svgscale\undefined%
  \makeatother%
  \begin{picture}(1,0.75223214)%
    \put(0,0){\includegraphics[width=\unitlength]{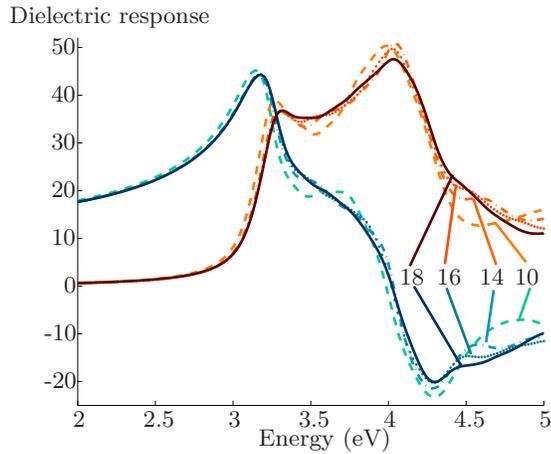}}%
    \put(0.12202411,0.05022321){\makebox(0,0)[lb]{\smash{2}}}%
    \put(0.23902455,0.05022321){\makebox(0,0)[lb]{\smash{2.5}}}%
    \put(0.38020759,0.05022321){\makebox(0,0)[lb]{\smash{3}}}%
    \put(0.49739509,0.05022321){\makebox(0,0)[lb]{\smash{3.5}}}%
    \put(0.63857812,0.05022321){\makebox(0,0)[lb]{\smash{4}}}%
    \put(0.75558036,0.05022321){\makebox(0,0)[lb]{\smash{4.5}}}%
    \put(0.89694866,0.05022321){\makebox(0,0)[lb]{\smash{5}}}%
    \put(0.07440446,0.11309554){\makebox(0,0)[lb]{\smash{$\text{-}20$}}}%
    \put(0.07440446,0.19252232){\makebox(0,0)[lb]{\smash{$\text{-}10$}}}%
    \put(0.10751518,0.27194866){\makebox(0,0)[lb]{\smash{0}}}%
    \put(0.08633214,0.35137723){\makebox(0,0)[lb]{\smash{10}}}%
    \put(0.08633214,0.4306183){\makebox(0,0)[lb]{\smash{20}}}%
    \put(0.08633214,0.51004464){\makebox(0,0)[lb]{\smash{30}}}%
    \put(0.08633214,0.58947098){\makebox(0,0)[lb]{\smash{40}}}%
    \put(0.08633214,0.66889955){\makebox(0,0)[lb]{\smash{50}}}%
    \put(0.4306183,0.01711339){\makebox(0,0)[lb]{\smash{Energy (eV)}}}%
    \put(0.85621205,0.28557321){\makebox(0,0)[lb]{\smash{10}}}%
    \put(0.79799107,0.28557321){\makebox(0,0)[lb]{\smash{14}}}%
    \put(0.6648058,0.28557321){\makebox(0,0)[lb]{\smash{18}}}%
    \put(0.72693527,0.28557321){\makebox(0,0)[lb]{\smash{16}}}%
%     \put(0.82756696,0.29057321){\makebox(0,0)[lb]{\smash{10}}}%
%     \put(0.76756696,0.29057321){\makebox(0,0)[lb]{\smash{14}}}%
%     \put(0.68756696,0.29057321){\makebox(0,0)[lb]{\smash{16}}}%
%     \put(0.62756696,0.29057321){\makebox(0,0)[lb]{\smash{18}}}%
    \put(0.01655536,0.72042411){\makebox(0,0)[lb]{\smash{Dielectric response}}}%
  \end{picture}%
\endgroup%
 \caption{(Color online). Convergence of $\varepsilon(\omega)$ (BSE) with respect to a traditional homogeneous sampling of the BZ. A shift $\textbf{s}$ in a non-symmetric direction is used (see text). 
 The number indicated is $n_k$ and the grid is therefore $(n_k\times n_k \times n_k | \textbf{s})$. The imaginary part is given in blue color while the real part is in orange-red. The full line corresponds to the finest grid that we have used, with $n_k$=18.
 Oscillations appear for energies larger than 3.2 eV, but are damped with increasing $n_k$. }
 \label{fig:convbse}
\end{figure}

\begin{figure}
\begin{minipage}[c]{\linewidth}
       \def\svgwidth{8cm}
\begingroup%
  \makeatletter%
  \providecommand\color[2][]{%
    \errmessage{(Inkscape) Color is used for the text in Inkscape, but the package 'color.sty' is not loaded}%
    \renewcommand\color[2][]{}%
  }%
  \providecommand\transparent[1]{%
    \errmessage{(Inkscape) Transparency is used (non-zero) for the text in Inkscape, but the package 'transparent.sty' is not loaded}%
    \renewcommand\transparent[1]{}%
  }%
  \providecommand\rotatebox[2]{#2}%
  \ifx\svgwidth\undefined%
    \setlength{\unitlength}{448bp}%
    \ifx\svgscale\undefined%
      \relax%
    \else%
      \setlength{\unitlength}{\unitlength * \real{\svgscale}}%
    \fi%
  \else%
    \setlength{\unitlength}{\svgwidth}%
  \fi%
  \global\let\svgwidth\undefined%
  \global\let\svgscale\undefined%
  \makeatother%
  \begin{picture}(1,0.75223214)%
    \put(0,0){\includegraphics[width=\unitlength]{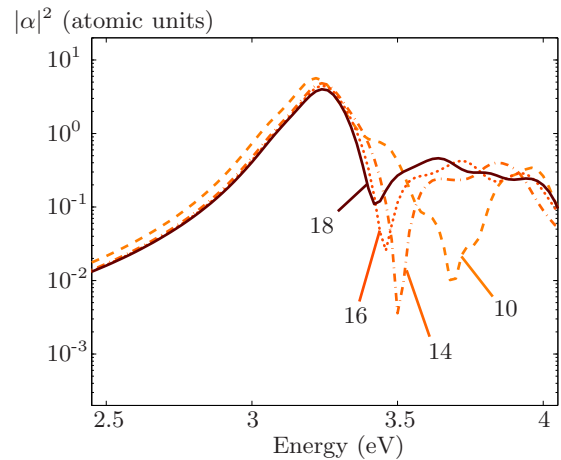}}%
    \put(0.13262679,0.04817679){\makebox(0,0)[lb]{\smash{2.5}}}%
    \put(0.38783482,0.04817679){\makebox(0,0)[lb]{\smash{3}}}%
    \put(0.61681473,0.04817679){\makebox(0,0)[lb]{\smash{3.5}}}%
    \put(0.87202455,0.04817679){\makebox(0,0)[lb]{\smash{4}}}%
    \put(0.05529018,0.15810982){\makebox(0,0)[lb]{\smash{$10^{\text{-}3}$}}}%
    \put(0.05529018,0.2803192){\makebox(0,0)[lb]{\smash{$10^{\text{-}2}$}}}%
    \put(0.05529018,0.40271652){\makebox(0,0)[lb]{\smash{$10^{\text{-}1}$}}}%
    \put(0.05529018,0.52492634){\makebox(0,0)[lb]{\smash{$10^{0}$}}}%
    \put(0.05529018,0.64732143){\makebox(0,0)[lb]{\smash{$10^{1}$}}}%
    \put(0.4306183,0.00506696){\makebox(0,0)[lb]{\smash{Energy (eV)}}}%
    \put(0.79724777,0.23357812){\makebox(0,0)[lb]{\smash{10}}}%
    \put(0.68895089,0.16139955){\makebox(0,0)[lb]{\smash{14}}}%
    \put(0.54821429,0.22147321){\makebox(0,0)[lb]{\smash{16}}}%
    \put(0.49227009,0.37183036){\makebox(0,0)[lb]{\smash{18}}}%
    \put(0,0.71197098){\makebox(0,0)[lb]{\smash{$|\alpha|^2$ (atomic units)}}}%
  \end{picture}%
\endgroup%
       \\
        (a)
   \end{minipage} \\
   \begin{minipage}[c]{\linewidth}
 \def\svgwidth{8cm}
\begingroup%
  \makeatletter%
  \providecommand\color[2][]{%
    \errmessage{(Inkscape) Color is used for the text in Inkscape, but the package 'color.sty' is not loaded}%
    \renewcommand\color[2][]{}%
  }%
  \providecommand\transparent[1]{%
    \errmessage{(Inkscape) Transparency is used (non-zero) for the text in Inkscape, but the package 'transparent.sty' is not loaded}%
    \renewcommand\transparent[1]{}%
  }%
  \providecommand\rotatebox[2]{#2}%
  \ifx\svgwidth\undefined%
    \setlength{\unitlength}{448bp}%
    \ifx\svgscale\undefined%
      \relax%
    \else%
      \setlength{\unitlength}{\unitlength * \real{\svgscale}}%
    \fi%
  \else%
    \setlength{\unitlength}{\svgwidth}%
  \fi%
  \global\let\svgwidth\undefined%
  \global\let\svgscale\undefined%
  \makeatother%
  \begin{picture}(1,0.75223214)%
    \put(0,0){\includegraphics[width=\unitlength]{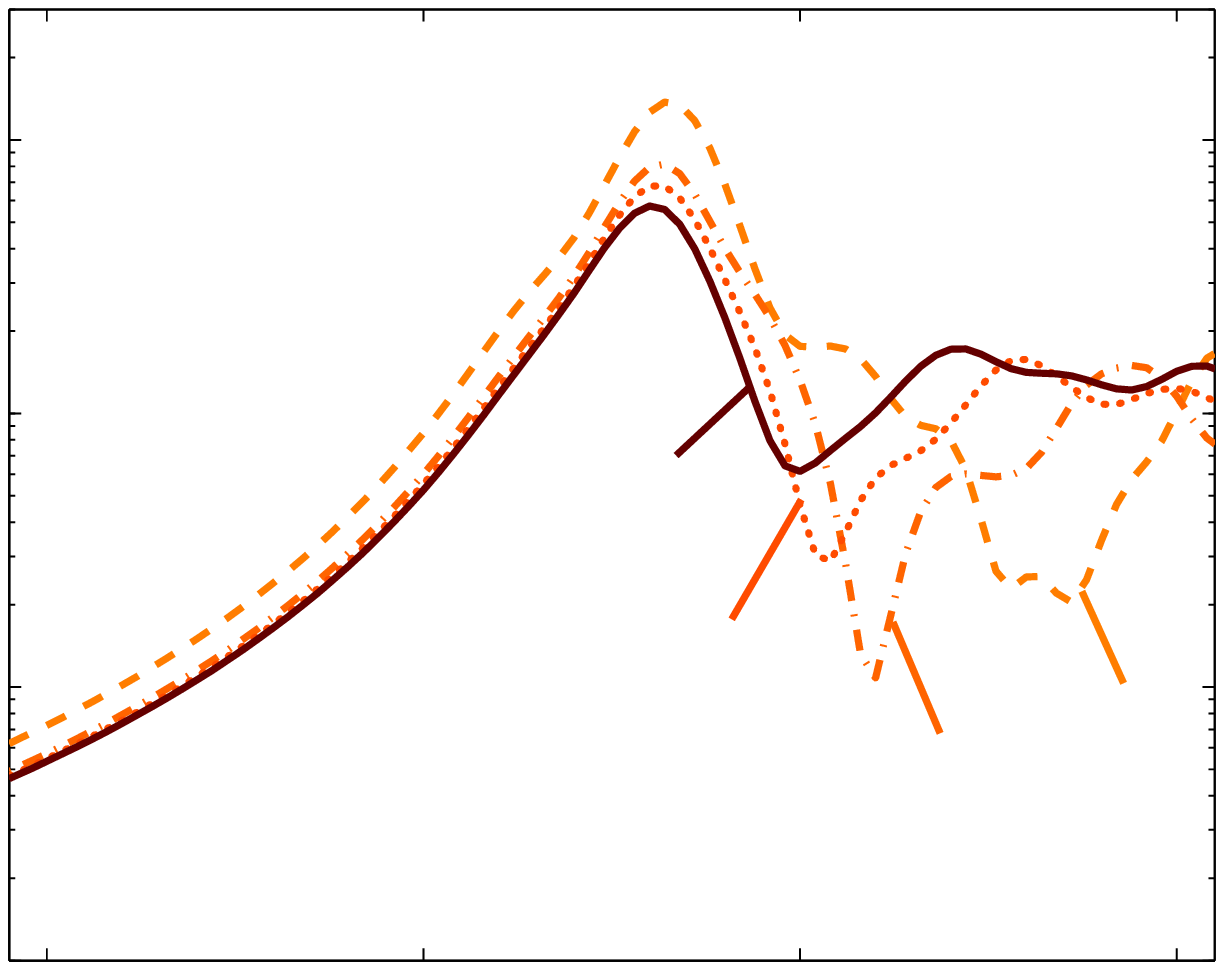}}%
    \put(0.13262679,0.04817679){\makebox(0,0)[lb]{\smash{2.5}}}%
    \put(0.38783482,0.04817679){\makebox(0,0)[lb]{\smash{3}}}%
    \put(0.61681473,0.04817679){\makebox(0,0)[lb]{\smash{3.5}}}%
    \put(0.87202455,0.04817679){\makebox(0,0)[lb]{\smash{4}}}%
    \put(0.05529018,0.07254464){\makebox(0,0)[lb]{\smash{$10^{\text{-}3}$}}}%
    \put(0.05529018,0.24851116){\makebox(0,0)[lb]{\smash{$10^{\text{-}2}$}}}%
    \put(0.05529018,0.42429241){\makebox(0,0)[lb]{\smash{$10^{\text{-}1}$}}}%
    \put(0.05529018,0.60007366){\makebox(0,0)[lb]{\smash{$10^{0}$}}}%
    \put(0.4306183,0.01506696){\makebox(0,0)[lb]{\smash{Energy (eV)}}}%
    \put(0.83724777,0.22357812){\makebox(0,0)[lb]{\smash{10}}}%
    \put(0.70895089,0.19139955){\makebox(0,0)[lb]{\smash{14}}}%
    \put(0.54821429,0.27147321){\makebox(0,0)[lb]{\smash{16}}}%
    \put(0.51227009,0.37183036){\makebox(0,0)[lb]{\smash{18}}}%
    \put(0,0.71197098){\makebox(0,0)[lb]{\smash{$|\alpha|^2$ (atomic units)}}}%
  \end{picture}%
\endgroup%
\\
   (b)
   \end{minipage}
 \caption{(Color online). Convergence of $|\alpha|^2$ with respect to a traditional homogeneous sampling of the BZ for (a) BSE and (b) RPA. A shift in a non-symmetric direction is used (see text). 
 The number indicated is $n_k$ and the grid is therefore $(n_k\times n_k \times n_k | \textbf{s})$. The convergence is difficult to achieve for energies larger than 3.2 eV.}
 \label{fig:convalpha}
\end{figure}

Fig. \ref{fig:convbse} presents the macroscopic dielectric function (from BSE) for different grids with increasing number of wavevectors, while Fig. \ref{fig:convalpha} presents the corresponding Raman intensity, both with excitonic contributions (BSE), and without excitonic contributions (RPA).
These grids, of size $N_k$, are shifted by the vector $\textbf{s} = (0.11,0.21,0.31)$ in reciprocal space along a non-symmetric direction.

As seen in Fig. \ref{fig:convbse}, for the computation of the macroscopic dielectric function, the oscillations present in the  $(10\times 10 \times 10 | \textbf{s})$ case are progressively damped when the density of the mesh is increased and a 
$(16\times 16 \times 16 | \textbf{s})$ grid gives converged results.

However, from Fig. \ref{fig:convalpha}, we observe that the convergence is much more difficult for the square of the Raman susceptibility, in the region beyond 3.2 eV (which corresponds to the optical gap). Important features still change
going from $(16\times 16 \times 16 | \textbf{s})$ to $(18\times 18 \times 18 | \textbf{s})$. Interestingly, 
such a difficult convergence is present both with and without 
excitonic contributions, as examplified by the upper and lower parts of Fig. \ref{fig:convalpha}.

With the method of shifted grids, convergence is not achievable, given our computational resources, beyond 3.2 eV. 
Indeed, the scaling of the method with respect to the sampling of the BZ is $\mathcal{O}(N_k^2)$, 
with $N_k$ the total number of k-points in the full BZ. With $N_k$=$18^3$, the convergence is not yet reached.

By contrast, convergence is much better below the gap value. 
In the next paragraphs, we first perform an analysis of the convergence for such frequencies, then turn to the higher-frequency part of the spectrum, for which we have developed a double-grid technique.

\subsection{\label{sec:results2}The convergence below the band gap}

In order to have a quantitative understanding of the convergence in the low-energy part, we use a Taylor expansion and give coefficients similar to the so-called Cauchy coefficients for the macroscopic dielectric function.\cite{Cauchy1836} 
Since the function is even with respect to the frequency, we can expand the absolute value of the Raman tensor and the real part of the dielectric function with even powers of the frequency:

\begin{align}
 \alpha(\omega) &= \alpha_0 + C^\alpha_2 \omega^2 + C^\alpha_4 \omega^4 + C^\alpha_6 \omega^6 \label{eq:cauchyalpha} \\
 \Re{\varepsilon(\omega)} &= \varepsilon_0 + C^\varepsilon_2 \omega^2 + C^\varepsilon_4 \omega^4 + C^\varepsilon_6 \omega^6 \label{eq:cauchyeps}
\end{align}

The coefficients can be obtained by a least-square fitting of the finite difference results until 1.5 eV
(see Tab.~\ref{tab:cauchy}).

\begin{table}
\begin{ruledtabular}
\begin{tabular}{c|cccc}
\textrm{$n_k$}&
\textrm{12}&
\textrm{14}&
\textrm{16}&
\textrm{18}\\
\colrule
$\varepsilon_0$ & $1.3276~10^{+1}$ & $1.3259~10^{+1}$ & $1.3256~10^{+1}$ &$1.3256~10^{+1}$ \\
$C^\varepsilon_2$& $7.7792~10^{-1}$ & $7.7571~10^{-1}$&  $7.7527~10^{-1}$&$7.7543~10^{-1}$\\
$C^\varepsilon_4$& $5.2340~10^{-2}$ &$5.2084~10^{-2}$ &$5.2020~10^{-2}$ &$5.2030~10^{-2}$ \\
$C^\varepsilon_6$& $5.8317~10^{-3}$&$5.7806~10^{-3}$&$5.7638~10^{-3}$&$5.7605~10^{-3}$\\
\hline
$\alpha_0$& $2.5167~10^{-2}$& $2.4464~10^{-2}$ &$2.4159~10^{-2}$ &$2.4027~10^{-2}$\\
$C^\alpha_2$& $4.8710~10^{-3}$& $4.7337~10^{-3}$& $4.6719~10^{-3}$ & $4.6460~10^{-3}$\\
$C^\alpha_4$& $5.7303~10^{-4}$ & $5.5503~10^{-4}$&$5.4689~10^{-4}$ & $5.4347~10^{-4}$\\
$C^\alpha_6$& $1.2462~10^{-4}$&  $1.1894~10^{-4}$&$1.1618~10^{-4}$&$1.1491~10^{-4}$
\end{tabular}
 \caption{Cauchy coefficients for $\varepsilon$ and for $\alpha$ within the BSE framework  (see Eqs. \eqref{eq:cauchyalpha} and \eqref{eq:cauchyeps}). The grids used are $(n_k \times n_k \times n_k | \textbf{s})$.}
 \label{tab:cauchy}
\end{ruledtabular}

\end{table}

\begin{figure}
 \def\svgwidth{8cm}
\begingroup%
  \makeatletter%
  \providecommand\color[2][]{%
    \errmessage{(Inkscape) Color is used for the text in Inkscape, but the package 'color.sty' is not loaded}%
    \renewcommand\color[2][]{}%
  }%
  \providecommand\transparent[1]{%
    \errmessage{(Inkscape) Transparency is used (non-zero) for the text in Inkscape, but the package 'transparent.sty' is not loaded}%
    \renewcommand\transparent[1]{}%
  }%
  \providecommand\rotatebox[2]{#2}%
  \ifx\svgwidth\undefined%
    \setlength{\unitlength}{448bp}%
    \ifx\svgscale\undefined%
      \relax%
    \else%
      \setlength{\unitlength}{\unitlength * \real{\svgscale}}%
    \fi%
  \else%
    \setlength{\unitlength}{\svgwidth}%
  \fi%
  \global\let\svgwidth\undefined%
  \global\let\svgscale\undefined%
  \makeatother%
  \begin{picture}(1,0.75223214)%
    \put(0,0){\includegraphics[width=\unitlength]{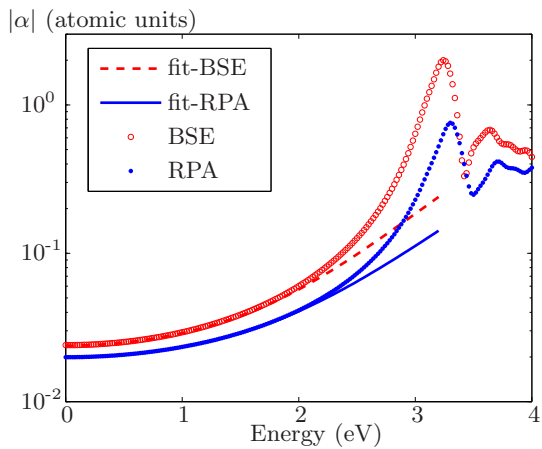}}%
    \put(0.12218734,0.05028101){\makebox(0,0)[lb]{\smash{0}}}%
    \put(0.31628059,0.05028101){\makebox(0,0)[lb]{\smash{1}}}%
    \put(0.51037342,0.05028101){\makebox(0,0)[lb]{\smash{2}}}%
    \put(0.70446624,0.05028101){\makebox(0,0)[lb]{\smash{3}}}%
    \put(0.89855907,0.05028101){\makebox(0,0)[lb]{\smash{4}}}%
    \put(0.05529018,0.07254464){\makebox(0,0)[lb]{\smash{$10^{\text{-}2}$}}}%
    \put(0.05529018,0.3193817){\makebox(0,0)[lb]{\smash{$10^{\text{-}1}$}}}%
    \put(0.05529018,0.56622098){\makebox(0,0)[lb]{\smash{$10^{0}$}}}%
    \put(0.43635654,0.01898734){\makebox(0,0)[lb]{\smash{Energy (eV)}}}%
    \put(0.29929241,0.62704687){\makebox(0,0)[lb]{\smash{fit-BSE}}}%
    \put(0.29929241,0.57012723){\makebox(0,0)[lb]{\smash{fit-RPA}}}%
    \put(0.29929241,0.51320759){\makebox(0,0)[lb]{\smash{BSE}}}%
    \put(0.29929241,0.45628795){\makebox(0,0)[lb]{\smash{RPA}}}%
    \put(0.03338228,0.70800211){\makebox(0,0)[lb]{\smash{$|\alpha|$ (atomic units)}}}%
  \end{picture}%
\endgroup%
 \caption{(Color online). Comparison between the absolute value of the Raman tensor obtained from the BSE and RPA, and their 
Cauchy polynomial fits in the low-frequency part of the spectrum. Coefficients have been obtained by fitting the data up to 1.5 eV for a uniformly shifted 18x18x18 k-grid.}
 \label{fig:fit-abs-alpha}
\end{figure}

The results of the fit obtained with this technique are presented in Fig. \ref{fig:fit-abs-alpha}. 
The range of validity of this fit goes beyond 2 eV. 
A fitting above 2 eV leads to an oscillatory behaviour in the very low energy range: the four-term expansion in Eq. \eqref{eq:cauchyalpha} and \eqref{eq:cauchyeps} is not accurate enough to describe the higher-energy part.

Cauchy coefficients are already well-converged for the $14^3$ grid (within a few percent for the first and second ones). 
However, such a fit does not correctly describe the resonance close to the gap energy.

\subsection{\label{sec:results3}The convergence above the band gap}

As mentioned earlier, we analyze the convergence of the Raman intensities in both the BSE case (Fig.~\ref{fig:convalpha} (a)) and in the RPA approximation (Fig.~\ref{fig:convalpha} (b)).
Both RPA and BSE present similar difficulties to converge the final results.
Hence, we can conclude that the convergence issue is {\it not} primarily due to the building up of the excitons that arises from the off-diagonal couplings, but is already present at the independent-particle level. 
Why the convergence rate in the case of Raman susceptibility is smaller than in the macroscopic dielectric function can be understood as follows.
The imaginary part of the dielectric function, for a given wavevector grid, is made of numerous broadened Dirac delta contributions, each of which corresponds to one transition from a valence band to a conduction band. 
In order for such spectrum to look smooth, the broadening should be comparable to the typical spacing between delta functions.
By contrast, the frequency-dependent Raman intensity is obtained by differentiating the dielectric function.
Hence, the Raman intensity evolution corresponds to the superposition of a large number of {\it derivatives} of broadened delta functions, whose oscillatory character are much stronger than the broadened Dirac functions. 
This is reflected at the level of the Raman intensity.

Having identified the problem, we design another strategy for sampling the BZ, that largely reduces the computational burden and memory requirements. 
In the same spirit as Ref.~\onlinecite{Kammerlander2012}, but with a rather different implementation, we introduce 
a double-grid technique. 

We perform a set of  BSE calculations, indexed by the label $i$, each with the same number of points in the BZ forming a ``coarse" grid,  differing by their shift $\textbf{s}_i$:

\begin{align}
 \left\{ \textbf{k} \right\}_i = (n_k \times n_k \times n_k | \textbf{s}_i) \label{eq:ki}
\end{align}
 
\begin{figure}
 \def\svgwidth{4cm}
\begingroup%
  \makeatletter%
  \providecommand\color[2][]{%
    \errmessage{(Inkscape) Color is used for the text in Inkscape, but the package 'color.sty' is not loaded}%
    \renewcommand\color[2][]{}%
  }%
  \providecommand\transparent[1]{%
    \errmessage{(Inkscape) Transparency is used (non-zero) for the text in Inkscape, but the package 'transparent.sty' is not loaded}%
    \renewcommand\transparent[1]{}%
  }%
  \providecommand\rotatebox[2]{#2}%
  \ifx\svgwidth\undefined%
    \setlength{\unitlength}{160.00012207bp}%
    \ifx\svgscale\undefined%
      \relax%
    \else%
      \setlength{\unitlength}{\unitlength * \real{\svgscale}}%
    \fi%
  \else%
    \setlength{\unitlength}{\svgwidth}%
  \fi%
  \global\let\svgwidth\undefined%
  \global\let\svgscale\undefined%
  \makeatother%
  \begin{picture}(1,1.00001961)%
    \put(0,0){\includegraphics[width=\unitlength]{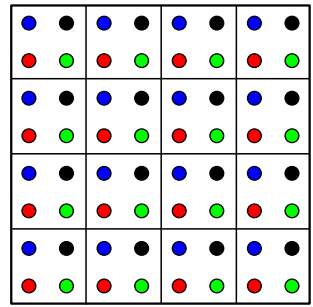}}%
  \end{picture}%
\endgroup%
 \caption{(Color online). Schematic representation of the sampling of the BZ by a double-grid technique in 2D, with $n_k = 4$ and $n_{div} = 2$ (see text for notations). One BSE computation is done on each grid (represented by four different colours in this picture). The final result is obtained by averaging the $n_{div}^2 = 4$ different computations.}
 \label{fig:interp-by-shifts}
\end{figure}

The shifts $\textbf{s}_i$ are chosen in order to obtain an homogeneous sampling of the subspace between the coarse points: 

\begin{align}
 \left\{ \textbf{s}_i \right\} = (n_{div} \times n_{div} \times n_{div} | \textbf{h})
\end{align}
with $n_{div}$ the number of subdivisions in each direction and $\textbf{h} = (1/2,1/2,1/2)$.
A 2D schematic representation is illustrated in Fig. \ref{fig:interp-by-shifts}.

With this technique, the macroscopic dielectric function is obtained by averaging the differents results computed on the ``coarse'' grids:

\begin{align}
 \varepsilon^{av}(\omega) = \frac{1}{n_{div}^3} \sum_i \varepsilon(\omega | \left\{\textbf{k} \right\}_i)
\end{align}
where $\varepsilon(\omega | \left\{\textbf{k} \right\}_i)$ is the macroscopic dielectric function obtained for the ``coarse'' computation with the grid $\left\{\textbf{k} \right\}_i$, Eq. \eqref{eq:ki}.

\begin{figure}
 \def\svgwidth{8cm}
\begingroup%
  \makeatletter%
  \providecommand\color[2][]{%
    \errmessage{(Inkscape) Color is used for the text in Inkscape, but the package 'color.sty' is not loaded}%
    \renewcommand\color[2][]{}%
  }%
  \providecommand\transparent[1]{%
    \errmessage{(Inkscape) Transparency is used (non-zero) for the text in Inkscape, but the package 'transparent.sty' is not loaded}%
    \renewcommand\transparent[1]{}%
  }%
  \providecommand\rotatebox[2]{#2}%
  \ifx\svgwidth\undefined%
    \setlength{\unitlength}{448bp}%
    \ifx\svgscale\undefined%
      \relax%
    \else%
      \setlength{\unitlength}{\unitlength * \real{\svgscale}}%
    \fi%
  \else%
    \setlength{\unitlength}{\svgwidth}%
  \fi%
  \global\let\svgwidth\undefined%
  \global\let\svgscale\undefined%
  \makeatother%
  \begin{picture}(1,0.75223214)%
    \put(0,0){\includegraphics[width=\unitlength]{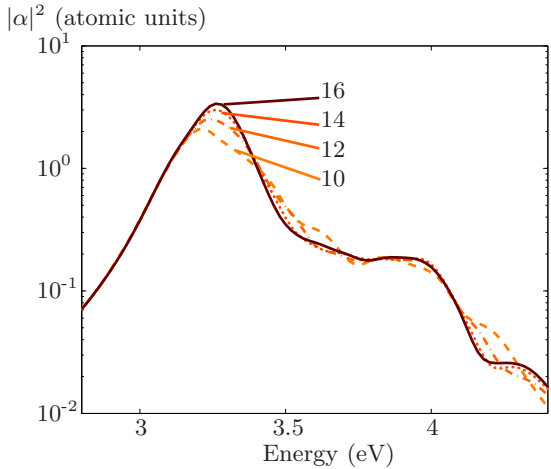}}%
    \put(0.21819196,0.04817679){\makebox(0,0)[lb]{\smash{3}}}%
    \put(0.44735938,0.04817679){\makebox(0,0)[lb]{\smash{3.5}}}%
    \put(0.70256696,0.04817679){\makebox(0,0)[lb]{\smash{4}}}%
    \put(0.05529018,0.07254464){\makebox(0,0)[lb]{\smash{$10^{\text{-}2}$}}}%
    \put(0.05529018,0.27641295){\makebox(0,0)[lb]{\smash{$10^{\text{-}1}$}}}%
    \put(0.05529018,0.48028348){\makebox(0,0)[lb]{\smash{$10^0$}}}%
    \put(0.05529018,0.68396652){\makebox(0,0)[lb]{\smash{$10^1$}}}%
    \put(0.4306183,0.00506696){\makebox(0,0)[lb]{\smash{Energy (eV)}}}%
    \put(0.52740402,0.60946429){\makebox(0,0)[lb]{\smash{16}}}%
    \put(0.52740402,0.56077455){\makebox(0,0)[lb]{\smash{14}}}%
    \put(0.52740402,0.51208259){\makebox(0,0)[lb]{\smash{12}}}%
    \put(0.52740402,0.46339286){\makebox(0,0)[lb]{\smash{10}}}%
    \put(0.00399554,0.7309375){\makebox(0,0)[lb]{\smash{$|\alpha|^2$ (atomic units)}}}%
  \end{picture}%
\endgroup%
 \caption{(Color online). Convergence of $|\alpha|^2$ with respect to the sampling of the Brillouin zone obtained with the double-grid technique. The number indicates $n_k$ for the ``coarse'' sampling of the Brillouin zone in each direction. 64 points are sampled in the subspace between the coarse points.}
 \label{fig:raman-interp}
\end{figure}

Fig.~\ref{fig:raman-interp} presents the results obtained with different coarse mesh samplings (different $n_k$), averaging over 64 calculations ($n_{div} = 4$ is kept constant). 
Of course, when $n_k$ becomes very large, the Raman spectrum must tend to the same spectrum as without this double-grid technique.
But the computational effort is largely reduced. 
Indeed, the residual fluctuation when going from $n_k=14$ to $n_k=16$ can be seen to be rather small already. 
In the RPA case, $n_k = 16$ with $n_{div} = 4$ corresponds exactly to a $(64 \times 64 \times 64 | (1/2,1/2,1/2) )$ uniform grid, that would be untractable in the BSE case. 
It is worth stressing that in the current method, we can take advantage of symmetries to reduce the number of ``coarse" grid calculations, since some meshes are equivalent. 
For example, the number of required computations with $n_{div} = 4$ falls down to 20 for the case where an atom is displaced and to 10 for the equilibrium position.

\section{Analysis of the theoretical results}

\begin{figure}
 \def\svgwidth{8cm}
\begingroup%
  \makeatletter%
  \providecommand\color[2][]{%
    \errmessage{(Inkscape) Color is used for the text in Inkscape, but the package 'color.sty' is not loaded}%
    \renewcommand\color[2][]{}%
  }%
  \providecommand\transparent[1]{%
    \errmessage{(Inkscape) Transparency is used (non-zero) for the text in Inkscape, but the package 'transparent.sty' is not loaded}%
    \renewcommand\transparent[1]{}%
  }%
  \providecommand\rotatebox[2]{#2}%
  \ifx\svgwidth\undefined%
    \setlength{\unitlength}{448bp}%
    \ifx\svgscale\undefined%
      \relax%
    \else%
      \setlength{\unitlength}{\unitlength * \real{\svgscale}}%
    \fi%
  \else%
    \setlength{\unitlength}{\svgwidth}%
  \fi%
  \global\let\svgwidth\undefined%
  \global\let\svgscale\undefined%
  \makeatother%
  \begin{picture}(1,0.8223214)%
    \put(0,0){\includegraphics[width=\unitlength]{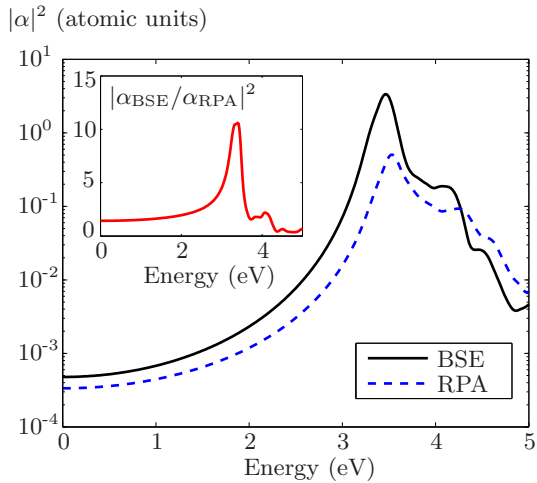}}%
    \put(0.12146607,0.04817679){\makebox(0,0)[lb]{\smash{0}}}%
    \put(0.27641295,0.04817679){\makebox(0,0)[lb]{\smash{1}}}%
    \put(0.43136161,0.04817679){\makebox(0,0)[lb]{\smash{2}}}%
    \put(0.58631027,0.04817679){\makebox(0,0)[lb]{\smash{3}}}%
    \put(0.7412567,0.04817679){\makebox(0,0)[lb]{\smash{4}}}%
    \put(0.89639062,0.04817679){\makebox(0,0)[lb]{\smash{5}}}%
    \put(0.05529018,0.07254464){\makebox(0,0)[lb]{\smash{$10^{\text{-}4}$}}}%
    \put(0.05529018,0.19494018){\makebox(0,0)[lb]{\smash{$10^{\text{-}3}$}}}%
    \put(0.05529018,0.31714955){\makebox(0,0)[lb]{\smash{$10^{\text{-}2}$}}}%
    \put(0.05529018,0.43954688){\makebox(0,0)[lb]{\smash{$10^{\text{-}1}$}}}%
    \put(0.05529018,0.5617567){\makebox(0,0)[lb]{\smash{$10^{0}$}}}%
    \put(0.05529018,0.68396652){\makebox(0,0)[lb]{\smash{$10^{1}$}}}%
    \put(0.4306183,0.00506696){\makebox(0,0)[lb]{\smash{Energy (eV)}}}%
    \put(0.75209821,0.18441964){\makebox(0,0)[lb]{\smash{BSE}}}%
    \put(0.75209821,0.14219464){\makebox(0,0)[lb]{\smash{RPA}}}%
    \put(0.1843375,0.36569866){\makebox(0,0)[lb]{\smash{0}}}%
    \put(0.31863839,0.36569866){\makebox(0,0)[lb]{\smash{2}}}%
    \put(0.45293973,0.36569866){\makebox(0,0)[lb]{\smash{4}}}%
    \put(0.16308214,0.40085491){\makebox(0,0)[lb]{\smash{0}}}%
    \put(0.16308214,0.4836317){\makebox(0,0)[lb]{\smash{5}}}%
    \put(0.14578304,0.56640625){\makebox(0,0)[lb]{\smash{10}}}%
    \put(0.14578304,0.6491808){\makebox(0,0)[lb]{\smash{15}}}%
    \put(0.26395089,0.32337723){\makebox(0,0)[lb]{\smash{Energy (eV) }}}%
    \put(0.03772321,0.75058036){\makebox(0,0)[lb]{\smash{$|\alpha|^2$ (atomic units)}}}%
    \put(0.20730625,0.62146607){\makebox(0,0)[lb]{\smash{$|\alpha_{\text{BSE}}/\alpha_{\text{RPA}}|^2$ }}}%
  \end{picture}%
\endgroup%
 \caption{(Color online). Comparison of the frequency evolution of $|\alpha|^2$ (in atomic units) with Bethe-Salpeter and RPA formalisms. 
 The inset shows the ratio between the two curves. The ratio goes up to 10 at the level of the direct band gap (3.2~eV).}
 \label{fig:compar-bse-rpa}
\end{figure}

In this section, we analyze in more details the importance of excitonic effects on the Raman spectrum.
A comparison between BSE and RPA results is reported in Fig. \ref{fig:compar-bse-rpa}.
Note how the excitonic effects amplify the Raman intensity by more than an order of magnitude in the band gap region.
Much smaller amplifications are observed for low frequencies.

Since the integral of the imaginary part of the dielectric
susceptibility is related to the plasmon frequency $\omega_p$ (the so-called f-sum rule): \cite{Johnson1974,Pines1963}

\begin{equation}
 \int_0^\infty \omega~\Im\{\varepsilon_{ij}(\omega)\} d\omega = \frac{1}{2} \pi \omega_p^2 \delta_{ij},
\end{equation}
and since $\omega_p$ does not depend on atomic positions, the integral of the imaginary part of the 
Raman susceptibility vanishes:

\begin{equation}
 \int_0^\infty \omega~\Im\{\alpha_{ij}(\omega)\} d\omega = 0. \label{eqn:intalpha}
\end{equation}

Accordingly, negative and positive regions are present in Fig. 7.
On the basis of Eq. \eqref{eqn:intalpha}, we can see that the difference between the BSE and RPA results for the Raman intensity is due to the lowering of the energy, and the amplification of the main peak of the imaginary part of the Raman susceptibility.

In the approximation for which the atomic displacement induces a global rigid shift of all the conduction bands with respect to all valence bands, with energy $\Delta \varepsilon=\varepsilon_{c\bf{k}}-\varepsilon_{v\bf{k}}$ (or, alternatively, if one transition dominates), the derivative with respect to an atomic displacement is related to the derivative with respect to frequency by:

\begin{equation}
\left. \frac{\partial \chi}{\partial \tau} \right|_\omega 
\approx 
\left. \frac{\partial  \chi}{\partial \Delta \varepsilon}\right|_\omega \frac{\partial \Delta \varepsilon}{\partial \tau} 
= 
\left. - \frac{\partial  \chi}{\partial \omega}\right|_\omega \frac{\partial \Delta \varepsilon}{\partial \tau}.
\end{equation}

This relation shows that the amplitude of the Raman effect will follow the variation of the band structure with the atomic displacement.\cite{Weber2000}

\begin{figure}
 \def\svgwidth{8cm}
\begingroup%
  \makeatletter%
  \providecommand\color[2][]{%
    \errmessage{(Inkscape) Color is used for the text in Inkscape, but the package 'color.sty' is not loaded}%
    \renewcommand\color[2][]{}%
  }%
  \providecommand\transparent[1]{%
    \errmessage{(Inkscape) Transparency is used (non-zero) for the text in Inkscape, but the package 'transparent.sty' is not loaded}%
    \renewcommand\transparent[1]{}%
  }%
  \providecommand\rotatebox[2]{#2}%
  \ifx\svgwidth\undefined%
    \setlength{\unitlength}{448bp}%
    \ifx\svgscale\undefined%
      \relax%
    \else%
      \setlength{\unitlength}{\unitlength * \real{\svgscale}}%
    \fi%
  \else%
    \setlength{\unitlength}{\svgwidth}%
  \fi%
  \global\let\svgwidth\undefined%
  \global\let\svgscale\undefined%
  \makeatother%
  \begin{picture}(1,0.75223214)%
    \put(0,0){\includegraphics[width=\unitlength]{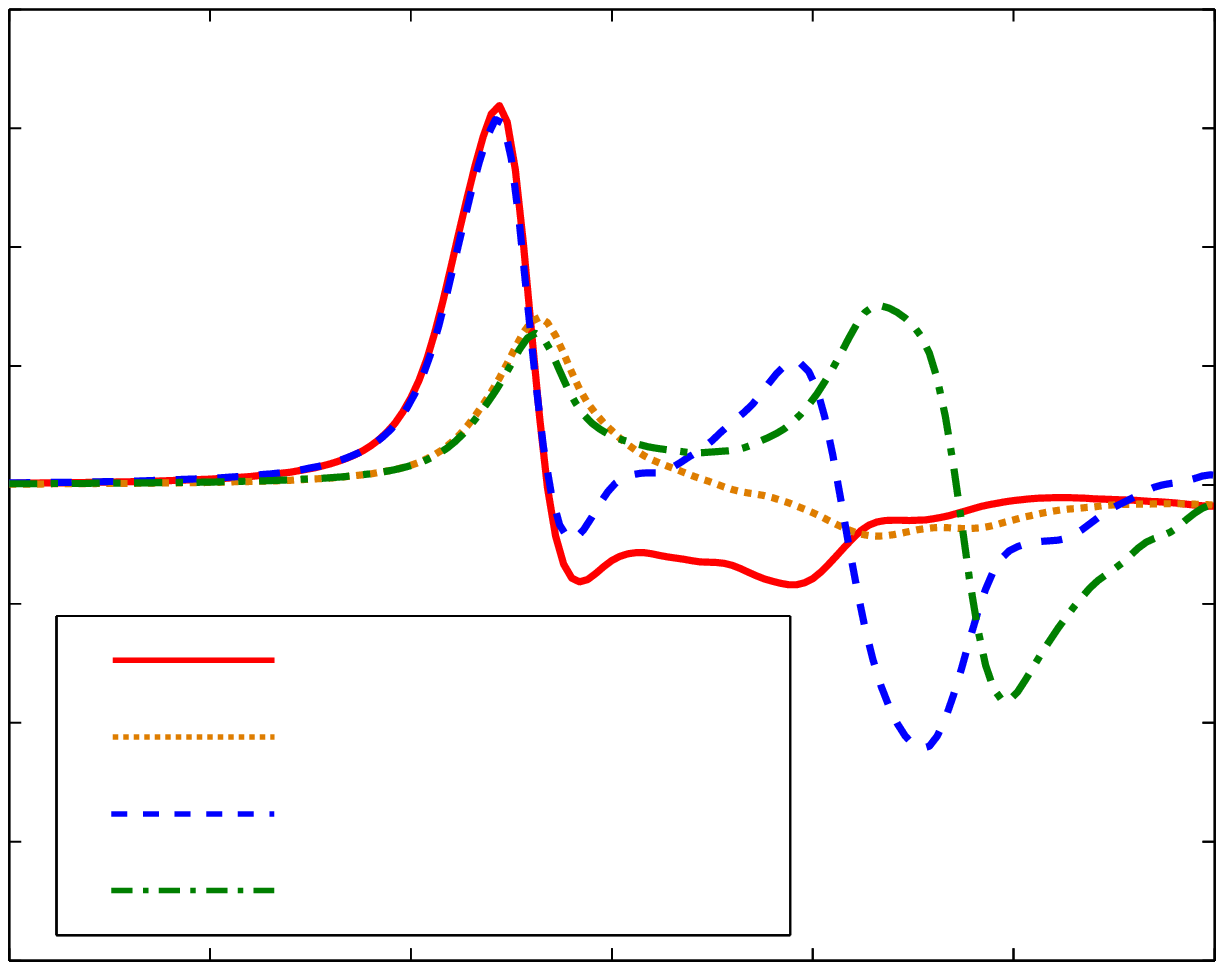}}%
    \put(0.12146607,0.04817679){\makebox(0,0)[lb]{\smash{2}}}%
    \put(0.23753795,0.04817679){\makebox(0,0)[lb]{\smash{2.5}}}%
    \put(0.37964955,0.04817679){\makebox(0,0)[lb]{\smash{3}}}%
    \put(0.49590848,0.04817679){\makebox(0,0)[lb]{\smash{3.5}}}%
    \put(0.63802009,0.04817679){\makebox(0,0)[lb]{\smash{4}}}%
    \put(0.75409152,0.04817679){\makebox(0,0)[lb]{\smash{4.5}}}%
    \put(0.89639062,0.04817679){\makebox(0,0)[lb]{\smash{5}}}%
   \put(0.92127232,0.07254464){\makebox(0,0)[lb]{\smash{-200}}}%
    \put(0.92127232,0.14899554){\makebox(0,0)[lb]{\smash{-150}}}%
    \put(0.92127232,0.22544643){\makebox(0,0)[lb]{\smash{-100}}}%
    \put(0.92127232,0.30189732){\makebox(0,0)[lb]{\smash{-50}}}%
    \put(0.92127232,0.37834821){\makebox(0,0)[lb]{\smash{0}}}%
    \put(0.92127232,0.45479911){\makebox(0,0)[lb]{\smash{50}}}%
    \put(0.92127232,0.53125){\makebox(0,0)[lb]{\smash{100}}}%
    \put(0.92127232,0.60770089){\makebox(0,0)[lb]{\smash{150}}}%
    \put(0.92127232,0.68396652){\makebox(0,0)[lb]{\smash{200}}}%
    \put(0.05517143,0.07254464){\makebox(0,0)[lb]{\smash{-2.0}}}%
    \put(0.05517143,0.14899554){\makebox(0,0)[lb]{\smash{-1.5}}}%
    \put(0.05517143,0.22544643){\makebox(0,0)[lb]{\smash{-1.0}}}%
    \put(0.05517143,0.30189732){\makebox(0,0)[lb]{\smash{-0.5}}}%
    \put(0.07017143,0.37834821){\makebox(0,0)[lb]{\smash{0.0}}}%
    \put(0.07017143,0.45479911){\makebox(0,0)[lb]{\smash{0.5}}}%
    \put(0.07017143,0.53125){\makebox(0,0)[lb]{\smash{1.0}}}%
    \put(0.07017143,0.60770089){\makebox(0,0)[lb]{\smash{1.5}}}%
    \put(0.07017143,0.68396652){\makebox(0,0)[lb]{\smash{2.0}}}%
    \put(0.4306183,0.01506696){\makebox(0,0)[lb]{\smash{Energy (eV)}}}%
    \put(0.33017857,0.26708304){\makebox(0,0)[lb]{\smash{$\text{Im}(\alpha_{BSE})$}}}%
    \put(0.33017857,0.21708304){\makebox(0,0)[lb]{\smash{$\text{Im}(\alpha_{RPA})$}}}%
    \put(0.33017857,0.17037232){\makebox(0,0)[lb]{\smash{$\text{Im}(d\varepsilon_{BSE}/d\omega)$}}}%
    \put(0.33017857,0.12037232){\makebox(0,0)[lb]{\smash{$\text{Im}(d\varepsilon_{RPA}/d\omega)$}}}%
    \put(0.03772321,0.75058036){\makebox(0,0)[lb]{\smash{$\text{Im}(\alpha)$ (atomic units)}}}%
    \put(0.73772321,0.75058036){\makebox(0,0)[lb]{\smash{$\text{Im}(d\varepsilon/d\omega)$ (eV$^{-1}$)}}}%
  \end{picture}%
\endgroup%
  \caption{(Color online). Comparison of the imaginary part of the Raman susceptibility and of the derivative of the susceptibility with 
 respect to frequency for the Bethe-Salpeter Equation (BSE) and for the Random-Phase Approximation (RPA). These computations are achieved with the double-grid technique for $n_k = 16$ and $n_{div}=4$.}
 \label{fig:compar-deriv}
\end{figure}

As represented in Fig.~\ref{fig:compar-deriv}, this approximation is only valid at the onset of absorption.
In this range of energies, the curves are qualitatively on top of each other.
This shows that the transition corresponding to the energy gap dominates in this range of energies. 
For higher energies, however, this approximation is not valid since each band can contribute differently from other bands in the Raman susceptibility.

\section{Comparison with experimental data}

\begin{figure}
  \def\svgwidth{8cm}

  \begingroup%
  \makeatletter%
  \providecommand\color[2][]{%
    \errmessage{(Inkscape) Color is used for the text in Inkscape, but the package 'color.sty' is not loaded}%
    \renewcommand\color[2][]{}%
  }%
  \providecommand\transparent[1]{%
    \errmessage{(Inkscape) Transparency is used (non-zero) for the text in Inkscape, but the package 'transparent.sty' is not loaded}%
    \renewcommand\transparent[1]{}%
  }%
  \providecommand\rotatebox[2]{#2}%
  \ifx\svgwidth\undefined%
    \setlength{\unitlength}{448bp}%
    \ifx\svgscale\undefined%
      \relax%
    \else%
      \setlength{\unitlength}{\unitlength * \real{\svgscale}}%
    \fi%
  \else%
    \setlength{\unitlength}{\svgwidth}%
  \fi%
  \global\let\svgwidth\undefined%
  \global\let\svgscale\undefined%
  \makeatother%
  \begin{picture}(1,0.75223214)%
    \put(0,0){\includegraphics[width=\unitlength]{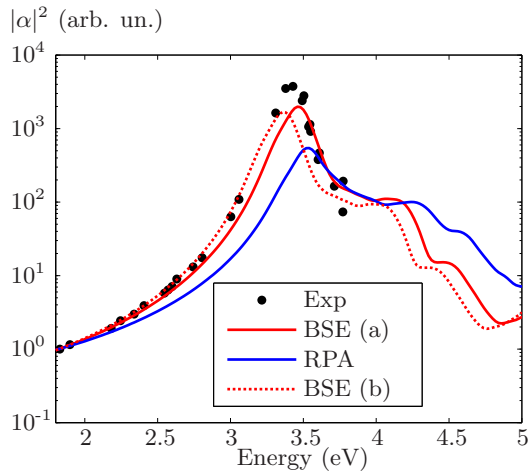}}%
    \put(0.16982857,0.04817679){\makebox(0,0)[lb]{\smash{2}}}%
    \put(0.27790179,0.04817679){\makebox(0,0)[lb]{\smash{2.5}}}%
    \put(0.41201562,0.04817679){\makebox(0,0)[lb]{\smash{3}}}%
    \put(0.52008929,0.04817679){\makebox(0,0)[lb]{\smash{3.5}}}%
    \put(0.65420312,0.04817679){\makebox(0,0)[lb]{\smash{4}}}%
    \put(0.76227679,0.04817679){\makebox(0,0)[lb]{\smash{4.5}}}%
    \put(0.89639062,0.04817679){\makebox(0,0)[lb]{\smash{5}}}%
    \put(0.05529018,0.07254464){\makebox(0,0)[lb]{\smash{$10^{\text{-}1}$}}}%
    \put(0.05529018,0.19494018){\makebox(0,0)[lb]{\smash{$10^0$}}}%
    \put(0.05529018,0.31714955){\makebox(0,0)[lb]{\smash{$10^1$}}}%
    \put(0.05529018,0.43954688){\makebox(0,0)[lb]{\smash{$10^2$}}}%
    \put(0.05529018,0.5617567){\makebox(0,0)[lb]{\smash{$10^3$}}}%
    \put(0.05529018,0.68396652){\makebox(0,0)[lb]{\smash{$10^4$}}}%
    \put(0.4306183,0.01506696){\makebox(0,0)[lb]{\smash{Energy (eV)}}}%
    \put(0.54147991,0.27660045){\makebox(0,0)[lb]{\smash{Exp}}}%
    \put(0.54147991,0.2273058){\makebox(0,0)[lb]{\smash{BSE (a)}}}%
    \put(0.54147991,0.17782768){\makebox(0,0)[lb]{\smash{RPA}}}%
    \put(0.54147991,0.12853393){\makebox(0,0)[lb]{\smash{BSE (b)}}}%
    \put(0.05252946,0.74304241){\makebox(0,0)[lb]{\smash{$|\alpha|^2$ (arb. un.)}}}%
  \end{picture}%
\endgroup%
  \caption{(Color online). Comparison of the theoretical and experimental Raman susceptibility of silicon.
  The experimental gap is reproduced by choosing a scissor value of 0.85 eV.
  A rigid energy shift of -0.1 eV is applied to results called ``BSE (a)'' in order to obtain ``BSE (b)''.
  The vertical values are given in arbitrary units in Ref.~\onlinecite{Compaan1984}. Therefore, these data are fitted to BSE and to RPA by means of a multiplicative factor fixed to pass through the first experimental point.}
  \label{fig:Ireduced}
\end{figure}

Fig. \ref{fig:Ireduced} shows the ab initio results for the Raman susceptibility of silicon, and compares these results with the experimental data obtained by Compaan and Trodahl,\cite{Compaan1984} who measured Raman intensity as a function of the frequency in silicon. 
In this figure, the theoretical results are obtained with a scissor value of 0.85~eV that reproduces the experimental gap at 0K (3.4~eV)
instead of 0.65~eV, which reproduces the theoretical GW gap (3.2~eV).

In terms of absolute value, the polarizability $a_{BSE}=19.75~\AA^2$ obtained at 1.1 eV compares reasonably well with the experimental data of $23\pm4~\AA^2$. 
The RPA value, $a_{RPA}=15.87~\AA^2$, does not match the experimental value. This confirms the need to correctly describe excitonic effects even for energies well below the gap.

We can distinguish three different regions with different behaviour: the low-energy (pre-resonance) region, from 2~eV to 3.2~eV, the band-gap region from 3.2~eV to 3.5~eV and the higher-energy region beyond 3.5~eV. 

\begin{figure}
 \def\svgwidth{8cm}
\begingroup%
  \makeatletter%
  \providecommand\color[2][]{%
    \errmessage{(Inkscape) Color is used for the text in Inkscape, but the package 'color.sty' is not loaded}%
    \renewcommand\color[2][]{}%
  }%
  \providecommand\transparent[1]{%
    \errmessage{(Inkscape) Transparency is used (non-zero) for the text in Inkscape, but the package 'transparent.sty' is not loaded}%
    \renewcommand\transparent[1]{}%
  }%
  \providecommand\rotatebox[2]{#2}%
  \ifx\svgwidth\undefined%
    \setlength{\unitlength}{448bp}%
    \ifx\svgscale\undefined%
      \relax%
    \else%
      \setlength{\unitlength}{\unitlength * \real{\svgscale}}%
    \fi%
  \else%
    \setlength{\unitlength}{\svgwidth}%
  \fi%
  \global\let\svgwidth\undefined%
  \global\let\svgscale\undefined%
  \makeatother%
  \begin{picture}(1,0.85223214)%
    \put(0,0){\includegraphics[width=\unitlength]{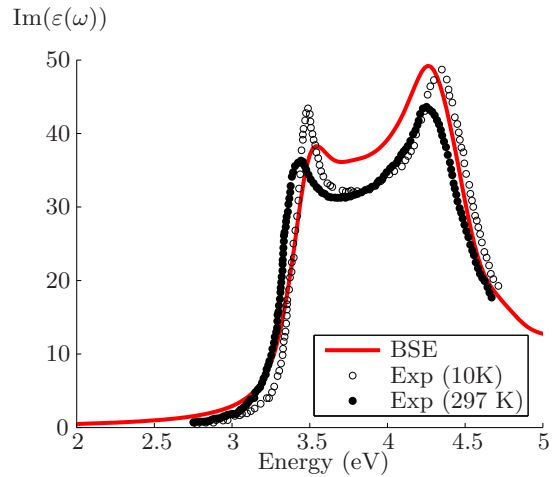}}%
    \put(0.12146607,0.04817679){\makebox(0,0)[lb]{\smash{2}}}%
    \put(0.23753795,0.04817679){\makebox(0,0)[lb]{\smash{2.5}}}%
    \put(0.37964955,0.04817679){\makebox(0,0)[lb]{\smash{3}}}%
    \put(0.49590848,0.04817679){\makebox(0,0)[lb]{\smash{3.5}}}%
    \put(0.63802009,0.04817679){\makebox(0,0)[lb]{\smash{4}}}%
    \put(0.75409152,0.04817679){\makebox(0,0)[lb]{\smash{4.5}}}%
    \put(0.89639062,0.04817679){\makebox(0,0)[lb]{\smash{5}}}%
    \put(0.0962125,0.07254464){\makebox(0,0)[lb]{\smash{0}}}%
    \put(0.07891339,0.19494018){\makebox(0,0)[lb]{\smash{10}}}%
    \put(0.07891339,0.31714955){\makebox(0,0)[lb]{\smash{20}}}%
    \put(0.07891339,0.43954688){\makebox(0,0)[lb]{\smash{30}}}%
    \put(0.07891339,0.5617567){\makebox(0,0)[lb]{\smash{40}}}%
    \put(0.07891339,0.68396652){\makebox(0,0)[lb]{\smash{50}}}%
    \put(0.4306183,0.01506696){\makebox(0,0)[lb]{\smash{Energy (eV)}}}%
    \put(0.65072991,0.20145089){\makebox(0,0)[lb]{\smash{BSE}}}%
    \put(0.65072991,0.15922589){\makebox(0,0)[lb]{\smash{Exp (10K)}}}%
    \put(0.65072991,0.1171875){\makebox(0,0)[lb]{\smash{Exp (297 K)}}}%
    \put(0.02399554,0.75502902){\makebox(0,0)[lb]{\smash{Im$(\varepsilon(\omega))$}}}%
  \end{picture}%
\endgroup%
 \caption{(Color online). Theoretical (BSE) and experimental absorption spectrum of silicon for 10 K and for 297 K.\cite{Jellison1983}}
 \label{fig:BSE+T}
\end{figure}

The Bethe-Salpeter method allows reproduction of the experimental amplification of the Raman susceptibility with the frequency. Agreement using this method is significantly better than the agreement obtained by RPA. 
In the band-gap region, however, there is a discrepancy between the theoretical and the experimental maximal Raman susceptibility.
The BSE maximum is nevertheless still closer to the experimental maximum than the RPA maximum.
% Interestingly, such a shift has already been discussed in the original paper\cite{Renucci1975} 
% and attributed to the inaccurate experimental data\cite{Philipp1972} used to post-process the Raman intensity.
Moreover, it is important to note that the BSE theoretical results obtained in this work are valid only at low temperature, whereas the experimental data are measured at 300 K.
The effect of the temperature on the absorption spectrum is illustrated in Fig. \ref{fig:BSE+T} where the first absorption peak at 10K is brought to lower energies at 297K, in agreement with the gap narrowing.
On the basis of this observation, we can expect an improvement in the agreement if temperature effects are included in the ab initio calculations.
As a first approximation, the effect of temperature on the gap energy can be mimicked by a rigid shift of the Raman intensity curve towards lower energies, as shown in Fig. \ref{fig:Ireduced} for data called ``BSE (b)''.
With this correction, the agreement is highly improved in the low-energy part of the Raman susceptibility. 
The amplification factor and post-resonance are however not significantly improved.
We attribute the disagreement to the different approximations we have performed so far, in particular within the BSE framework and the quasi-static approximation.

Our ab initio approach is not able to describe the higher-energy region as well as the lower-energy region.
Without the temperature correction, the BSE theoretical results underestimate the evolution of intensities in the pre-resonance region.
The inclusion of temperature leads to a partial improvement in the overall agreement, except for the post-resonance region.
The slope of decrease in the post-resonance region is in agreement with the experimental slope while the intensity value is underestimated.

The present approach relies on several approximations, whose roles still need to be analyzed. 
We have neglected, among others, the phonon frequency in the ``quasi-static'' approach, the quadratic response with respect to atomic displacements, the self-consistency with the $GW$ approximation, 
the non-Hermitian coupling within BSE, the frequency dependence of the BSE hamiltonian, 
indirect transitions, the additional effects due to phonons (the experimental data have been obtained at room temperature).
The latter effects could be included using a method similar to the method presented by Marini et al.\cite{Marini2008}
Calculations including all these effects would require computational resources unavailable to us at present.
 
Despite these effects, in all regions, the BSE results are in better agreement with experimental data than the RPA results. This clearly indicates the importance of excitonic effects for an accurate ab initio description of Raman spectra.

Note also that silicon might possess specific characteristics that induce the good agreement here observed with the present method and associated approximations. 
Such a good agreement might not be observed for materials with different characteristics, like a lower crystalline symmetry, 
the presence of stronger spin-orbit coupling, the presence of some ionicity, the presence of multiple types of atoms, etc ...

\section{Conclusions and perspectives}

A technique for the ab initio study of the resonant Raman intensity has been proposed, and applied to silicon.
The technique relies on finite differences of the macroscopic dielectric function evaluated for distorted geometries,
and includes excitonic effects by solving the Bethe-Salpeter equation.
We found that convergence of the Raman spectrum with respect to the number of wavevectors
 used to sample the Brillouin zone is problematic. To tackle this problem, we proposed a double-grid averaging process
that significantly improves convergence while keeping computational effort at a reasonable level.

The Bethe-Salpeter results are in better agreement with experimental results than those results obtained without excitonic effects (RPA). 
For laser energies in the band-gap region or lower, the agreement is excellent if one takes 
into account the small rigid shift that is needed to align the first peak of the theoretical dielectric 
absorption, as well as the experimental shift for the same temperature as the Raman spectrum.
The agreement is still not perfect however 
in the high-energy region. This may be attributed to many different effects, to be examined in future works.

\begin{acknowledgments}
Y.G. and M.G. wish to acknowledge the financial support of the Fonds National de la Recherche Scientifique (FNRS, Belgium).

The authors would like to thank Yann Pouillon and Jean-Michel Beuken for their valuable technical
support and help with the test and build system of ABINIT, and Nicola Thrupp for her careful reading of the final version of the article.

Computational resources have been provided by the supercomputing facilities of
the Universit\'e catholique de Louvain (CISM/UCL) and the Consortium des
Equipements de Calcul Intensif en F\'ed\'eration Wallonie Bruxelles (CECI) funded by
the Fonds de la Recherche Scientifique de Belgique (FRS-FNRS).
\end{acknowledgments}

\bibliography{raman3_final_fsum.bib}

\end{document}